# Fiber-Amplified Pulses Beyond the Gain Narrowing Limit for Seeding an OPA

**BRITTANY LU,**[1*] **KEITH WERNSING,**[2*] **SERGIO CARBAJO**[1,3,4,*]

[1]*Department of Electrical and Computer Engineering, University of California Los Angeles, Los Angeles, CA 90095, USA*
[2]*Mesa Photonics, 1550 Pacheco St, Santa Fe, NM 87505, USA*
[3]*Department of Physics and Astronomy, University of California Los Angeles, Los Angeles, CA 90095, USA*
[4]*SLAC National Accelerator Laboratory and Stanford University, 2575 Sand Hill Road, Menlo Park, CA, USA*
*\*Corresponding authors: scarbajo@ucla.edu, kwernsing@mesaphotonics.com, brittanylu@ucla.edu*

**We report on the design and numerical simulation of a tunable, near-infrared (NIR) femtosecond noncollinear optical parametric amplifier (OPA) seeded by a nonlinear fiber amplifier operating in the parabolic and gain-managed regime. With microjoule-level pump pulses in the visible, we achieved amplification bandwidths of signal and idler pulses between 1010-1170 nm and 920-1050 nm, respectively, and whose second harmonics correspond to the visible range between ~460-585 nm. The estimated quantum efficiency for the conversion of the pump to signal and idler pulses combined was greater than 15%. This appreciable value in a single OPA stage is due to the high spectral energy density of the fiber-amplified broadband pulse - about 30 times greater in the 1 μm range when compared to white-light continuum (WLC) pulses, which are often used to seed commercial OPAs. When combined with their excellent spatial beam quality and full coherence across their entire bandwidth, fiber-amplified pulses enable more efficient engineering and application of OPAs in the visible and NIR.**

## 1. INTRODUCTION

Optical parametric amplifiers (OPA) are powerful tools that produce light of variable wavelengths and ultrashort pulse durations from an otherwise fixed-wavelength laser source [1,2] Wavelength tunability is important, for example, in spectroscopy, where light is used to excite and probe the different energy transitions of a sample of interest. OPAs in the visible and near-infrared (NIR) have been employed for imaging deep tissues through multi-photon microscopy [3–5], for imaging blood flow through *in vivo* tracking of harmonic particles in living animals [6], for probing ultrafast exciton and phonon dynamics in quantum materials [7], and for studying charge photogeneration and recombination in solar cells via time-resolved transient absorption spectroscopy measurements [8]. Improving every amplification stage is paramount to visible and NIR OPAs.

The OPA process begins with the generation of the initial signal, namely the *seed*. The generation scheme and attributes of the seed dictate many of the performance parameters of the OPA. One such important factor is the number of seed photons. A higher number of photons allows for efficient saturation, especially in the noncollinear OPA (NOPA) configuration, translating to higher output energy stabilities [9,10]. Too low of a photon count in the spectral region to be amplified could induce significant shot-to-shot energy fluctuations [10]. Most commercial ultrafast OPAs use a white light generated (WLG) seed that spectrally extends from the visible to the NIR. Such a pulse is produced by focusing an intense ultrashort seed pulse inside a bulk laser host material. Through the filamentation process stemming from multiple nonlinear effects such as self-focusing, multiphoton ionization, self-phase modulation (SPM), and self-steepening [1,11], the spectral range of the initial pulse is stretched. However, these multiple nonlinear processes contribute to OPA noise. Though WLG continuum can exhibit an appreciable pulse-to-pulse energy stability [9,12], typical spectral energy densities of the WLG continuum pulse are limited, e.g. 10-30 pJ/nm in the visible and NIR range for white-light generated in YAG and sapphire crystals [9]. In the near ultraviolet (UV) range, using $CaF_2$ disks, higher pulse energies were obtained and used to seed a NOPA [10], though these disks need to be continuously moved to avoid cumulative photodamage [13]. Further, at high input energies, filamentation results in the emission of a chromatic conical ring [1,12,14–16], requiring additional spatial filtering to clean the spatial beam profile.

Despite its broad bandwidth, the light in the spectral region surrounding the initial seed is incoherent and unsuitable for usage, especially in an OPA [9,17]. In this region, due to SPM, the chirp of the pulse greatly increases, causing the spectrum to be highly complex as opposed to being nearly flat [9,17,18]. This significantly reduces the energy of a WLG OPA seed as a large portion of the original white light seed energy is contained in this spectral region [1].

While the bulk of earlier WLG studies were based on Ti:sapphire laser systems, Yb-lasers show emerging potential for efficient WLG in the visible [19], and mid-IR [20], albeit still exhibiting low pulse energy [19]. When combined with energetic pump pulses generated from Yb-based laser systems, OPAs can scale dramatically in output energy [1] and can deliver significant amplification bandwidths in the noncollinear geometry [21] or at degeneracy [22], as well as providing high gain in a few mm crystal [23]. The small quantum defect of Yb atoms, when pumped at 940 or 980 nm and lasing at 1030 nm, reduces thermal problems [24,25]. Compared to Ti:sapphire laser sources, which have relatively limited average powers, Yb-doped materials can handle higher average power operations, permitting repetition rates above tens of kHz and higher energy stability. This is especially desirable in experiments that require averaging over multiple shots, high acquisition speeds, and/or high average power but low energies per pulse.

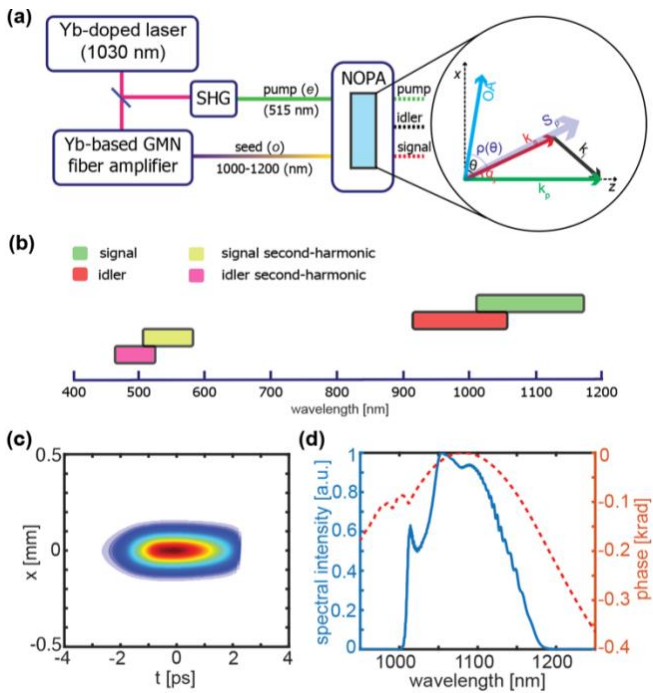

Fig. 1. (a) Simplified schematic. The second harmonic generation (SHG) from a Yb-doped laser (1030 nm) is used to produce the OPA pump (*e*-polarized). The OPA seed is obtained from the output of the Yb-fiber amplifier (*o*-polarized). Inset: The pump wavevector $\mathbf{k}_p$ (green arrow) is parallel to the *z*-axis and makes an angle θ with the crystal optic axis (OA, blue arrow). The pump Poynting vector $\mathbf{S}_p$ is at an angle ρ(θ) with respect to the OA. The signal wavevector $\mathbf{k}_s$ (red arrow) points toward $\mathbf{S}_p$ at the pump-tilt angle $α_s$. The idler wavevector $\mathbf{k}_i$ (black arrow) satisfies $\mathbf{k}_p = \mathbf{k}_s + \mathbf{k}_i$ at perfect phase-matching. (b) Schematic of the OPA signal (green) and idler (red) wavelength ranges and their second harmonics. (c) Spatiotemporal profile of the seed pulse, centered at *t* = 0 and *x* = 0. (d) Spectral intensity (black solid) and phase (red dashed) of the fiber-amplified pulse.

As an alternative to WLG, we propose using a broadband OPA seed source with high spectral energy density that can be generated directly from a 1030 nm pulse derived from a high-power, high repetition rate Yb-laser. By exploiting a notable advancement in nonlinear fiber amplification, i.e., gain-managed nonlinear (GMN) amplification in Yb-doped fibers [26,27], we demonstrate that GMNs substantially support the production of broadband pulses with spectral energy densities 20 times higher that WLG, thus reducing filamentation-induced white light noise in OPAs. Unlike WLG, the resulting smoothly broadened spectra surround the initial 1030 nm fiber-amplified seed pulse. In addition, fiber pulses offer high-quality and uniform spatiotemporal mode profiles, in contrast to the chromatic ring structure observed from filamentation [10].

In this study, we present numerical simulations of a tunable NOPA pumped by the second harmonic of a Yb-laser system and seeded by fiber-amplified pulses generated inside a Yb-fiber amplifier (Fig. 1.a). We obtained signal and idler pulses ranging from 1026 to 1144 nm, and 937 to 1035 nm, respectively, for which the second harmonics correspond to the visible (Fig. 1b). The resulting signal pulses exhibit excellent spatiotemporal beam qualities suitable for application-based use, further staged amplification, or for pulse shaping. More notably, in a single OPA stage, we achieve quantum efficiencies estimated to be greater than 16%, with signal and idler pulse energies each around 0.6 µJ when pumped by 3 µJ 515 nm light. Such a high conversion efficiency is made possible by the increased spectral energy density of the fiber-amplified seed.

## 2. FIBER PULSE AMPLIFICATION

Recent results have shown that double-clad Yb-gain fiber amplifiers can generate chirped picosecond pulses with bandwidths supporting sub-30 fs compressed pulse durations, with pulse energies up to the microjoule level [26,27]. These pulses are generated by controlled nonlinear amplification acting in concert with a red-shifting gain spectrum. The launched seed pulses initially amplify in the self-similar regime, growing in bandwidth, temporal duration and energy while maintaining a parabolic temporal profile and a linear chirp. As the co-propagating pump light gradually depletes, the quasi-three-level nature of the Yb-gain spectrum causes the center of the gain bandwidth to red-shift. The pulse no longer amplifies self-similarly, but in this gain-managed regime, the amplification still contains several important qualities – a smooth temporal envelope, linear chirp, and spectral broadening with a spectral profile that lacks the extreme undulations of typical SPM-dominated nonlinear propagation [26,27]. These properties enable the energetic output pulse to immediately provide a high-quality seed for an OPA.

To physically model nonlinear pulse propagation in the fiber amplifier, we used the generalized nonlinear Schrödinger equation (GNLSE), which models the evolution of the complex electric field envelope as it propagates along the fiber length [28,29]. The GNLSE was numerically implemented with the split-step propagator method, where nonlinearities such as SPM, self-steepening, and the stimulated Raman effect are handled in the temporal domain, while dispersion and gain are applied linearly in the spectral domain. Gain was modeled using the rate equations for a two-level system [30]. Wavelength-dependent absorption and emission cross-sections were used to compute the gain spectrum at each step along the gain fiber length.

The fiber chain consisted of three components – a 2m long passive segment, a 3m long double-clad gain fiber, and a 15cm passive fiber. The passive fibers were also double-clad so that 976nm pump laser light could co-propagate in the multimode core. A 10/125 core/cladding geometry was used to simulate Liekki's Yb1200-10/125DC-PM gain fiber and the matching passive fiber P-10/125DC-PM. A 0.5nJ Gaussian pulse centered at 1030 nm with a 4.5nm spectral width is launched into the fiber chain and is reshaped into a parabolic profile through SPM in the first fiber section. Upon entering the gain fiber, it initially undergoes self-similar (or shape-preserving) amplification, where the combined interplay of gain, SPM, and dispersion allow the pulse to grow in energy, bandwidth, and duration while maintaining its parabolic profile [27,31]. The nonlinear amplification then moves into the gain-managed regime, characterized by a red-shift in the gain spectrum. Finally, a short piece of passive fiber terminated with a fiber connector generates additional spectral broadening. Compared to SPM-generated bandwidths in passive fiber, the output from the GMN amplifier exhibits a relatively smooth spectral envelope between 1020 and 1170nm. The simulated pulse spatiotemporal profile and spectrum after exiting the fiber chain are shown in Fig. 1(b) and 1(c), respectively. The pulse has a total energy of 152 nJ and a nominal spectral energy density of about 600 pJ/nm across the middle of the spectrum, which is nominally 30 times greater than that of WLG pulses.

## 3. SIMULATION RESULTS AND DISCUSSION

We perform simulations in Chi2D, a nonlinear pulse propagation code that implements a 2+1-dimensional – in ($x$, $z$) space and time – Fourier split-step algorithm, and accounts for dispersion, diffraction, and walk-off in birefringent crystals [32]. Chi2D allows for arbitrary pulse profile inputs by loading their spectra and phase data. The program outputs the complex electric field amplitudes of the ordinary and extraordinary fields. The nonlinear material considered was a 1-mm long Type-I β-barium borate (BBO), which is suitable for high average power operations because of its superior thermal properties and small linear absorption coefficients ($10^{-4}$-$10^{-2}$ cm$^{-1}$ in the ~0.5-1 μm wavelength region) [33,34]. All geometrical values presented here were selected to produce signal pulses that have a full-width-half-maximum (FWHM) spectral bandwidth of around 14-17 nm (~120-1340 cm$^{-1}$) at ~1.02-1.15 μm, which can be easily further stretched or compressed, depending on the requirements. The crystal length was chosen based on commercial availability and considering with an upper boundary that too long of a crystal can result in temporal and spatial walk-off due to the different group velocities of all pulses in the nonlinear configuration. The length of BBO is less than the maximum pulse splitting length for the signal and pump. The effective nonlinear optical coefficient was set to 2 pm/V, consistent with reported data [2]; this is a comparably high value, allowing for high single-pass gain [33]. The wavevector of the extraordinarily ($e$) polarized pump propagates along the $z$-axis, at the phase-matching angle, θ, relative to the optic axis (OA) (Fig. 1.a). The pump is modeled as a 200-fs transform-limited Gaussian pulse having 3 μJ total energy and 515 nm center wavelength, which can be produced from the second harmonic generation (SHG) of commercially available 6-W-level Yb-laser amplifiers at 1 MHz repetition rate, assuming 50% SHG conversion efficiency. The output spectral and phase data from the fiber pulse amplification propagator described above was input as the seed pulse in Chi2D. Both pulses were assumed to have Gaussian spatial profiles with 1/$e^2$ radii of 0.11 mm, corresponding to pump peak intensity of 74 GW/cm$^2$. The signal range covers the degeneracy wavelength of 1030 nm at which both signal and idler photons emerge. The spatial separation of the two $o$-polarized pulses was achieved by employing a non-collinear geometry, with the initial seed propagating at an angle known as the pump-tilt angle, $α_s$, relative to the pump. Note that we refer to the signal pulse as the pulse that propagates along the same direction of the initial seed, and the idler as the generated pulse.

Wavelength tunability was accomplished by varying the pump-seed group delay, $τ_{GD}$, which adjusts the overlap in time between the pump and the different spectral regions of broadband seed, separated because of temporal chirp (Fig. 2a). To ensure near-perfect phase-matching with the wavelength to be amplified, we simultaneously adjusted the phase-matching angle θ. By increasing the pump group delay from -1400 to 1500 fs, we amplified signal wavelengths between 1026 and 1144 nm. Outside of this range, sidebands appear in the signal spectrum. Phase-matching angles for fixed $α_s$ can be found in SNLO [35], which was further tuned to improve the quality of the spatiotemporal profile of the signal pulse. For the group delay range given above, the phase-matching angle decreased from θ=25.17° to θ=24.72°. Variations in the phase-matching angle by ±0.2° did not significantly affect the signal spectra or spatiotemporal profiles. The corresponding pump walk-off angles are between 3.43° and 3.48°. To increase the spatial overlap between the pump and the signal Poynting vectors inside the crystal, we fixed the pump-signal angle to be $α_s$ = 3.5°, close to the spatial walk-off angles of the pump.

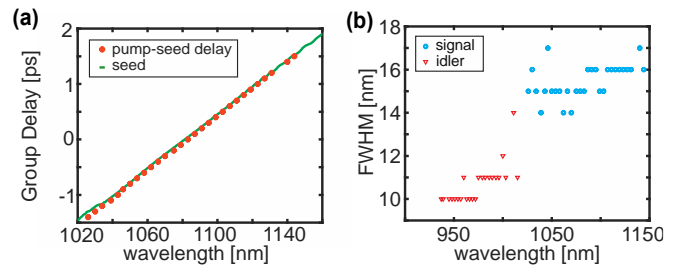

Fig. 2. (a) Seed group delay (solid blue) and applied pump-seed delay (red circle) as a function of wavelengths and the wavelengths amplified, respectively. (b) FWHM spectral bandwidth of the resulting signal (blue circle) and idler (red triangle) pulses.

The resulting signal pulse bandwidths at FWHM ranged from 14 nm at 1039 nm to 17 nm at 1144 nm (Fig. 2b, blue circles). The shortest pulse duration at FWHM was 140 fs at the center wavelength of 1026 nm, and the longest duration was 163 fs at 1087 nm. The time bandwidth products of the pulses were 0.558 at 1026 nm and 0.666 at 1087 nm, more than the bandwidth-limited value of 0.44, assuming a Gaussian temporal pulse profile, which can be caused by the angular dispersion of the signal. For the idler pulses, the bandwidths were smaller, the smallest being 10 nm between 937 and 957 nm wavelength, and the largest, 14 nm for at 1011 nm wavelength (Fig. 2.b, red triangles).

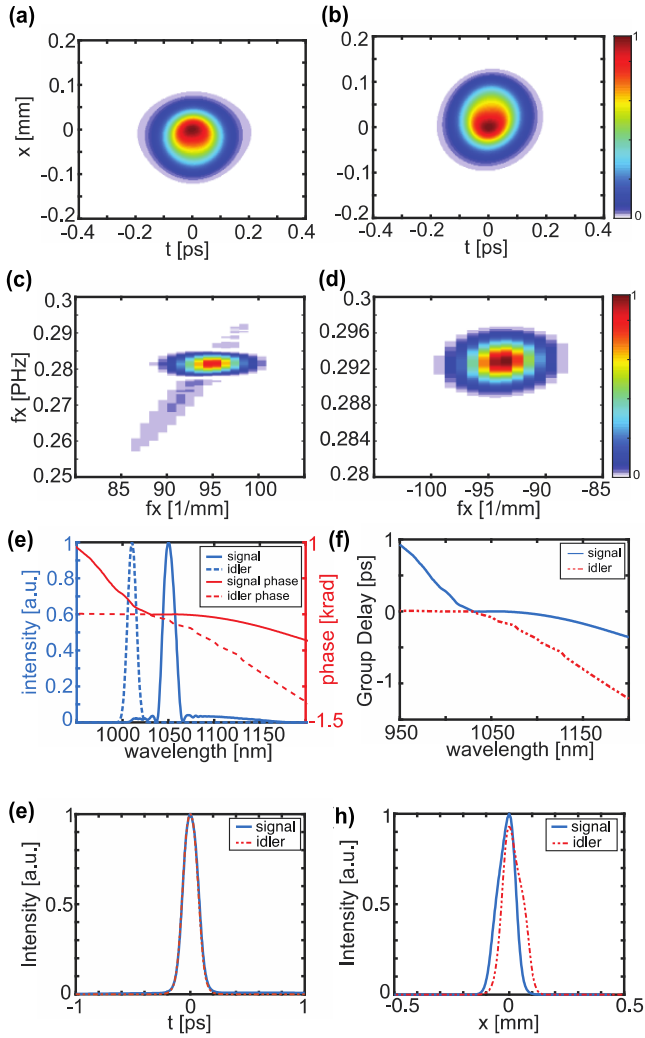

Fig. 3. Spatiotemporal profile of the (a) signal and (b) idler pulses. Time $t = 0$ and position $x = 0$ are defined at the center of the pulses. Intensity profile of the (c) signal and (d) idler pulses in the inverse spatial ($f_x$) and spectral ($f_t$) domains. (e) Spectral intensity of the signal (red) and idler (blue) and their corresponding phases (dashed); normalized individually at 3.5° and -3.35°, respectively. (f) Group delay of the signal (dashed blue) and idler (solid red). (g) Normalized temporal profile of the signal (dashed blue) and idler (solid red). (h) Spatial profile of the signal (dashed blue) and idler (solid red). Results given for $\tau_{GD}$ = -800 fs and $\theta$ =25.1°.

An example of the resulting pulses for $\tau_{GD}$ = -800 fs and $\theta$ =25.1° which amplified a center wavelength of 1050 nm, is shown in Fig 3.a. The spatiotemporal profiles of the signal and idler are shown in Fig. 3.a and b, respectively. The slopes of intensity profiles in different domains reflect first-order distortions [36]. The pulse front tilt angle is characterized by the slope of the profile in the ($x$, $t$) domain at the centers of the spatial and temporal profile. The resulting signal pulses exhibit no pulse-front tilt, but the idler pulses have a slight pulse-front tilt of around 3°. Pulse front tilt can also arise from angular dispersion. Typically, in a NOPA setup, the idler presents a large spectral angular dispersion [2], but here the signal is more angularly dispersed than the idler. As can be seen from the slopes of the intensity profiles in the ($f_x$, $f_t$) domains (Fig. 3.c and d) and the spectral intensity profile (Fig. 3.e, solid blue), there are residual spectral components from the signal between 1075-1200 nm. The corresponding spectral intensity and phase, and group delays are shown in Fig. 3.e and f, respectively. Slices across $x$ and $t$ at the center of gravity of the pulse are shown in Fig. 3.g and h. Both the signal and idler pulses have a nearly Gaussian temporal profile, but the spatial profile along $x$ appears to be skewed, especially for the idler pulse.

Assuming that the transverse radii in the pump at the end of the crystal length are equal, typical energies transferred to both the signal and the idler from the pump was slightly greater than 30% of the initial pump energy for all cases studied, which corresponds to signal and idler pulses around 0.5 μJ. The ratio between the input pump photons and the output signal photons is a good measure of amplification efficiency; in our case, the calculations render a ~16% conversion of pump photons to signal and idler photons combined.

## 3. CONCLUSIONS

We presented numerical simulations of a fiber-amplified parabolic pulse used for seeding an OPA that is based on commercially available 6-W-level Yb-laser amplifiers operating at around 1 MHz repetition rate. With 3 μJ pump pulse energies at 515 nm, approximately 16% of pump photons were converted to signal and idler photons combined. By virtue of the high spectral energy density and smoothly broadband shape of our seed pulse, we achieve high conversion efficiency in a single amplification stage when pumping at sub-microjoule pulse energies. The spectral bandwidths of the signal and idler pulses were less than ~140 cm$^{-1}$ and the average pulse duration was 156 fs. We were able to achieve broad signal and idler wavelength tunability between 1026-1170 nm and 920-1051 nm, respectively. To meet the demands of applications requiring higher pulse energies and power, the resulting pulses can be further amplified in subsequent OPA stages. Further, to extend to the visible spectral range, the amplified signal and idler pulses can undergo second-harmonic generation, where the wavelengths span 505-585 nm and 460-525 nm, respectively. However, at higher optical powers, OPA performance degrades due to absorption within the nonlinear crystal. Therefore, it is important to mitigate the effects of absorption-based heating, which could limit attainable power, spectral bandwidth, and beam quality [33]. When combined with proper thermal management techniques and design, the resulting energetic pulses produced from a tunable Yb-based OPA system seeded by a broadband, spectrally dense GMN-amplified parabolic pulse could have numerous uses in ultrafast spectroscopy applications that favor high repetition rates and high energy pulses consistent across the visible and near-IR; for example, in neuroscience, biophysics, solid-state physics, and photovoltaics, where Yb-based OPA systems are currently being employed.

**Funding.** We acknowledge funding from the U.S. Department of Energy Office of Sciences under contracts DE-SC0022559, DE-AC02-76SF00515, DE-SC0022464

**Acknowledgments.** We'd like to thank Huseyin Cankaya, Anne-Laure Calendron for fruitful discussions and feedback, and Randy Lemons for his assistance with simulations.

**Disclosures**. "The authors declare no conflicts of interest."

**Data availability.** Data underlying the results presented in this paper are not publicly available at this time but may be obtained from the authors upon reasonable request.